# Hygro-mechanical properties of paper fibrous networks through asymptotic homogenization and comparison with idealized models.

E. Bosco[a,b,∗], R.H.J. Peerlings[b], M.G.D. Geers[b]

[a]*Materials innovation institute (M2i), P.O. Box 5008, 2600 GA Delft, The Netherlands*
[b]*Department of Mechanical Engineering, Eindhoven University of Technology, P.O. Box 513, 5600 MB Eindhoven, The Netherlands*

## Abstract

This paper presents a multi-scale approach to predict the effective hygro-mechanical behaviour of paper sheets based on the properties of the underlying fibrous network. Despite the vast amount of literature on paper hygro-expansion, the functional dependence of the effective material properties on the micro-structural features remains yet unclear. In this work, a micro-structural model of the paper fibrous network is first developed by random deposition of the fibres within a planar region according to an orientation probability density function. Asymptotic homogenization is used to determine its effective properties numerically. Alternatively, two much more idealized micro-structural models are considered, one based on a periodic lattice structure with a regular network of perpendicular fibres and one based on the Voigt average. Despite their simplicity, they reproduce representative micro-structural features, such as the orientation anisotropy and network level hygro-elastic properties. These alternative models can be solved analytically, providing closed-form expressions that explicitly reveal the influence of the individual micro-scale parameters on the effective hygro-mechanical response. The trend predicted by the random network model is captured reasonably well by the two idealized models. The resulting hygro-mechanical properties are finally compared with experimental data reported in the literature, revealing an adequate quantitative agreement.[1]

## 1. Introduction

The dimensional stability of paper is a well known problem that affects several kinds of engineering applications, ranging from printing to packaging and storage operations. The response of paper to humidity variations is due to complex mechanisms originating at the level of the individual fibres [1]. The moisture induced swelling of a single fibre is transferred to the entire network through the inter-fibre bonds, which play a crucial role in determining the macroscopic response. At this scale, paper may exhibit different types of in-plane or out-of plane deformations [2]. A key challenge is to understand how to bridge phenomena at the fibre and network level to the effective behaviour of paper, in order to formulate scale transition relations that allow to predict the effective response as a function of the properties of the underlying micro-structure.

---

The hygro-expansion of paper has been thoroughly investigated in the literature, mostly through experiments at the macroscopic scale, see e.g. [3, 4, 5, 6, 7, 8, 9]. In these studies, the influence of several features of the fibrous network on the effective response has been analysed, for instance the inter-fibre bonds properties [7, 8]. Despite these valuable studies, an explicit, predictive relation between the micro-structural parameters, such as the hygroscopic and mechanical properties of the single fibre and the network structural characteristics, and macro-scale paper hygro-expansion has not yet been established.

A modelling approach can offer a substantial step forward in this direction. Classical homogenization provides the hygro-expansive coefficients of a general multi-phase material based on the hygro-mechanical properties of its constituting phases [10]. The resulting relations are, however, formulated with respect to an arbitrary geometry, and their specification for a network of fibres is not trivial. In [11], a general formula for the hygro-expansive coefficients of paper is obtained as a function of the longitudinal and transverse hygro-expansions of the single fibres. These are related by weights incorporating the role of the hygro-mechanical interactions within the inter-fibre bonds of the network. Whereas the proposed formula can explain qualitatively the trend of some experimental results [5, 11], it does not yet provide a quantitative estimate of the effective hygro-expansion properties for a specific network geometry.

The main goal of this work is to propose a multi-scale approach to calculate the effective hygro-mechanical properties of paper. This is done by developing a model of the underlying fibrous network and by studying its overall response through homogenization, by which the influence of the individual micro-structural parameters can be captured. In principle, different hypotheses can be made in formulating the model, which to some extent influence the results. In this paper, a dual strategy is followed. A detailed description of the network level *and* two highly idealized micro-structural models are developed and compared. As common assumptions, a two dimensional representation and perfectly bonded fibres are considered.

The first model is based on a periodic representative volume element (RVE) of the fibrous network, in which straight fibres are randomly positioned in the plane and follow a certain orientation distribution. This kind of description has analogies in the literature [12, 13, 14]; the main originality here is the fact that the fibres are modelled as two dimensional domains, characterized by both longitudinal and transverse hygro-mechanical properties. The effective hygro-mechanical properties of the network are computed via asymptotic homogenization ([15, 16, 17, 18]) following the general procedure developed in [19]. The method mainly consists in solving numerically a set of boundary value problems in relation to a given network geometry. This provides the displacement fluctuations at the network level due to mechanical and hygroscopic loads, which are needed to compute the effective elastic moduli and hygro-expansion coefficients. In the present manuscript, this methodology is specifically applied to paper. This includes the estimate of effective material response using appropriate network properties, a comparison with experimental data and a study on the influence of the different material parameters within specific ranges typical for paper.

Alternatively, two idealized micro-structural models are proposed. The first one simplifies paper to a lattice structure, consisting of a unit-cell with elements in two orthogonal directions [20, 21]. The effect of the fibre orientation is included by defining weighted thicknesses for the fibres in the two directions, whose ratio is obtained by integration of the orientation distribution function. A key aspect of this approach is that, like the full network model, it distinguishes between the free-standing, unbonded fibre segments and the inter-fibre bonds. The model is solved analytically through homogenization, allowing to recover the hygro-mechanical response as an explicit function of the micro-structural properties. The second idealized description is based on the Voigt estimate, for which paper is considered a composite plate with an infinite number of layers oriented according to the probability density function.

A comparative study between the proposed methodologies is finally performed, to define their range of applicability and to identify their practical use for several industrially relevant applications.

This paper is organized as follows. In Section 2, the network model and the calculation of the effective hygro-mechanical material properties through asymptotic homogenization are proposed. The idealized descriptions of the paper fibrous network are detailed in Section 3. The comparison between the proposed approaches is made in Section 4, in terms of the resulting effective hygro-mechanical properties. Conclusions are finally given in Section 5.

## 2. Fibrous network model and asymptotic homogenization

The detailed fibrous network model and the asymptotic homogenization procedure to calculate the effective hygro-mechanical properties of paper have been developed in Reference [19]. The main concepts are summarized here for the sake of clarity. Paper is approximated as a two dimensional material subjected to in-plane loads. A global reference system $(x, y)$ is defined, which is aligned with respect to the material principal directions, the machine direction and the cross direction. The position vector reads $\mathbf{x} = x\mathbf{e}_x + y\mathbf{e}_y$, with $\mathbf{e}_i, (i = x, y)$ the unit vectors of a Cartesian vector basis.

### 2.1. Network geometry

The micro-structural fibrous network is described through the periodic repetition of a two dimensional unit-cell. Note that real paper micro-structures are not periodic. However, the unit-cell is assumed to be a representative volume element (RVE), which is sufficiently large to contain enough micro-structural information and to represent a real, stochastic paper network [22]. The RVE is assumed to be square, with edge $L$ and area $Q : (0, L) \times (0, L)$. The geometry of each fibre is defined to be rectangular, with length $l$ and width $w$. A rectangular cross section $w \times t$ is assumed, with $t$ the thickness of the fibre. The fibres are positioned according to a uniform random point field, which determines the location of their geometrical centres $\mathbf{x}_c$ in the domain $Q$. Each fibre is characterized by an angle $\theta$ that follows a wrapped Cauchy orientation distribution probability density function [23]:

$$f(\theta) = \frac{1}{\pi} \frac{1 - q^2}{1 + q^2 - 2q\cos(2\theta)} \qquad (1)$$

where $-\pi/2 < \theta \leq \pi/2$ is the angle between the fibre axis and the $x-$direction (machine direction) and $0 \leq q < 1$ is a measure of the anisotropy of the network orientation. The portions of the fibres that fall outside the box $(0, L) \times (0, L)$ along a certain edge are trimmed and copied into the volume at the opposite edge. This ensures the periodicity of the RVE.

The number of fibres $n_f$ of the network is characterized by the areal coverage $\bar{c}$,

$$\bar{c} = \frac{n_f wl}{Q} = \frac{b_p}{b_f} \qquad (2)$$

defined as the ratio between the total area of the fibres and the area of the volume element. The last equality in (2) relates the coverage to the basis weight (i.e. the ratio between mass and area) of the fibres, $b_f$, and of the network, $b_p$; it will be used in the comparison with the idealized model of the network. Note that the coverage represents the average number of fibre layers characterizing the network and thus defines its average thickness to be $t\bar{c}$. During the network generation procedure, the fibres are deposited in the domain $Q$. Since the model is two dimensional, it is not possible to take into account the porosity of the system. The order of deposition is not considered, and all the fibres that overlap in a certain point of the domain are considered to be perfectly bonded to each other. Essentially, the thickness direction is collapsed into a plane, keeping track however, locally, of the number of fibres through the thickness and their orientations. The information on the average thickness provided by the coverage will be used to compute the local material properties, as discussed in Section 2.2.

Each fibre is treated as a two dimensional object. Generating a consistent finite element mesh for the complex network geometry may be cumbersome, especially in the bonds, in which two or more fibres with different orientations overlap. For this reason, a simplified procedure is adopted. A dense regular grid of square finite elements of edge length $l_e = w/\xi$ ($\xi \geq 1$, integer) is defined over the RVE. A certain finite element $e$ is considered to be part of a fibre if its geometrical center is located inside the fibre domain, i.e. the rectangular region of area $l \times w$ centred at $\mathbf{x}_c$ and oriented at the angle $\theta$. This implies that the fibre boundaries will reflect a jagged shape. This may affect the local stress and strain distributions. However, if $\xi$ is sufficiently large, the resulting effective hygro-mechanical properties are only minimally influenced.

Figure 1 presents an example of two networks of coverage $\bar{c} = 0.5$, with unit-cell edge $L$, fibre length $l = L/2$ and aspect ratio $l/w = 20$. Note that in Figure 1 this aspect ratio is considerably smaller than that used in what follows ($l/w = 100$) in order to better visualise the finite element discretization. Figure 1(a) shows a uniform fibre orientation distribution ($q = 0$), Figure 1(b) illustrates an anisotropic fibre orientation distribution, with $q = 0.5$. A detail of the finite element mesh of network (b) is shown in Figure 1(c).

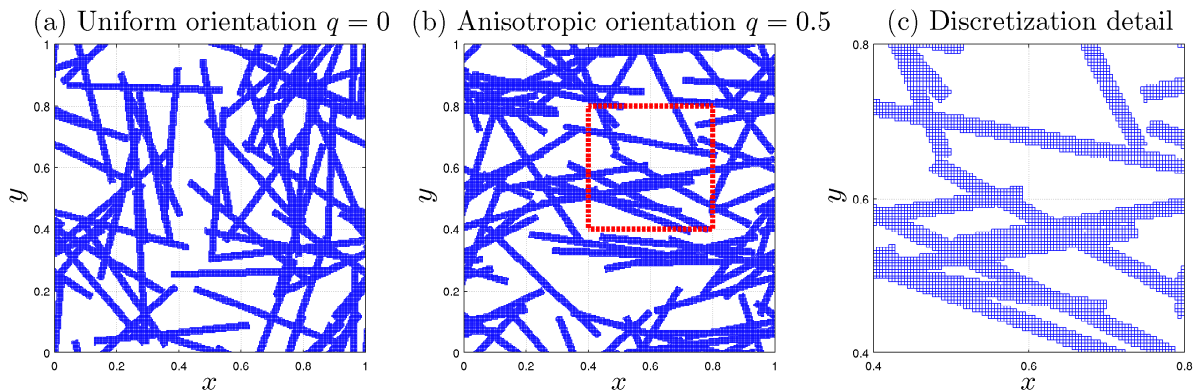

Figure 1: Example of two network configurations of coverage $\bar{c} = 0.5$, with (a) a uniform orientation distribution ($q = 0$) and (b) an anisotropic distribution ($q = 0.5$). A detail of the finite element discretization of network (b) is given in (c). The values of coverage and fibre aspect ratio ($l/w = 20$) are quite low for real paper; they are used here solely for a clearer illustration.

*2.2. Constitutive model*

Several works in the literature (see e.g. [24, 25, 26]) assume paper fibres as transversely isotropic materials. In a plane stress state, the hygro-elastic constitutive response of a single fibre is defined through its transversely isotropic fourth order elasticity tensor $^4\mathbf{C}^f$ and second order hygro-expansion tensor $\boldsymbol{\beta}^f$. In Voigt notation, they can be expressed as a matrix and a column as follows:

$$\underline{C}^f = \begin{pmatrix} \frac{E_\ell}{(1-\nu_{\ell t}\nu_{t\ell})} & \frac{\nu_{t\ell}E_\ell}{(1-\nu_{\ell t}\nu_{t\ell})} & 0 \\ \frac{\nu_{\ell t}E_t}{(1-\nu_{\ell t}\nu_{t\ell})} & \frac{E_t}{(1-\nu_{\ell t}\nu_{t\ell})} & 0 \\ 0 & 0 & G_{\ell t} \end{pmatrix}; \quad \underset{\sim}{\beta}^f = \begin{pmatrix} \beta_\ell \\ \beta_t \\ 0 \end{pmatrix} \quad (3)$$

The following definitions are employed in (3): $E_\ell$ and $E_t$ are the longitudinal and transverse moduli of elasticity, respectively; $G_{\ell t}$ is the in-plane shear modulus; the in-plane Poisson's ratios are $\nu_{\ell t}$ and $\nu_{t\ell}$, with $E_\ell \nu_{t\ell} = E_t \nu_{\ell t}$; $\beta_\ell$ and $\beta_t$ indicate the longitudinal and transverse hygro-expansion coefficients of the fibres, respectively. Note that the constitutive properties (3) are relative to a local coordinate system defined along the principal directions of a fibre oriented at an angle $\theta$. To extract the effective properties of the fibrous network, quantities

(3) have to be expressed with respect to the global reference system $(x, y)$ by performing a rotation for each of the fibres in the network. The obtained properties of the $k-th$ fibre depend on its orientation $\theta^{(k)}$ and will be indicated from now on as $^4\mathbf{C}^{f(k)}$ and $\boldsymbol{\beta}^{f(k)}$. The adopted procedure is rather standard; for details on the derivation of the expressions of $^4\mathbf{C}^{f(k)}$ and $\boldsymbol{\beta}^{f(k)}$, the reader may refer to [21].

The elasticity tensor $^4\mathbf{C}(\mathbf{x})$ and the hygro-expansion tensor $\boldsymbol{\beta}(\mathbf{x})$ characterizing the local response of the fibrous network are finally calculated by considering each point $\mathbf{x}$ of the plane as a parallel arrangement of fibres, each of them with thickness $t$. The stiffnesses and the hygro-expansion contributions of the individual layers are added and the result is divided by the nominal thickness $t\bar{c}$,

$$^4\mathbf{C}(\mathbf{x}) = \frac{1}{t\bar{c}} \sum_{k=1}^{n_b} t\, ^4\mathbf{C}^{f(k)}(\mathbf{x}) \;; \quad \boldsymbol{\beta}(\mathbf{x}) = \frac{1}{t\bar{c}} (^4\mathbf{C}(\mathbf{x}))^{-1} : \sum_{k=1}^{n_b} t \left( ^4\mathbf{C}^{f(k)}(\mathbf{x}) : \boldsymbol{\beta}^{f(k)}(\mathbf{x}) \right) \quad (4)$$

where $n_b$ is the number of fibres passing through the considered point. The operation $^4\mathbf{A} : \mathbf{B} = A_{ijhk}B_{kh}$ defines the inner product between tensors, in which Einstein's summation convention is used. Relation (4) corresponds, for $n_b \geq 2$, to the assumption of full strain compatibility (i.e. perfect bonding) between the fibre layers that compose the bonds. This allows to capture the interplay between hygro-expansion and mechanics due to the different longitudinal and transverse properties of fibres, which strongly influences the effective hygro-mechanical behaviour of the fibrous network.

*2.3. Effective hygro-mechanical properties through asymptotic homogenization*

The effective hygro-mechanical behaviour of the networks generated in Section 2.1 is estimated by means of asymptotic homogenization [15, 16, 17], which assumes a strong separation of scales, i.e. the length over which macro-scale loads vary should be much larger than the characteristic size of the micro-scale, related to the fibre length or RVE size. The asymptotic homogenization framework allows to transform the original, local constitutive relation characterized by the rapidly oscillating hygro-mechanical properties $^4\mathbf{C}(\mathbf{x})$ and $\boldsymbol{\beta}(\mathbf{x})$ into two series of mathematical problems. The first serves the calculation of the first order influence functions $^3\mathbf{N}^1(\mathbf{x}) \in H^1(Q)$ and $\mathbf{b}^1(\mathbf{x}) \in H^1(Q)$, which are the periodic solutions of the so-called cell-problems. Tensor $^3\mathbf{N}^1(\mathbf{x})$ is symmetric in two of the three indices, i.e. $^3N^1_{ijk} = {}^3N^1_{ikj}$. The latter are two boundary value problems defined on the representative volume element and depending only on the micro-structural constitutive properties $^4\mathbf{C}(\mathbf{x})$ and $\boldsymbol{\beta}(\mathbf{x})$:

$$\boldsymbol{\nabla} \cdot (^4\mathbf{C}(\mathbf{x}) : (\boldsymbol{\nabla}^3\mathbf{N}^1(\mathbf{x}) + {}^4\mathbf{I}^S)) = {}^3\mathbf{0} \quad (5)$$

$$\boldsymbol{\nabla} \cdot (^4\mathbf{C}(\mathbf{x}) : (\boldsymbol{\nabla}\mathbf{b}^1(\mathbf{x}) - \boldsymbol{\beta}(\mathbf{x}))) = \mathbf{0} \quad (6)$$

with $\boldsymbol{\nabla}\cdot$ indicating the divergence operator, $\boldsymbol{\nabla}$ the gradient operator and $^4\mathbf{I}^S$ the fourth-order symmetric identity tensor, defined as $I^S_{ijkl} = (\delta_{il}\delta_{jk} + \delta_{ik}\delta_{jl})/2$. Equations (5) and (6) are completed with periodic boundary conditions and the requirement that $^3\mathbf{N}^1(\mathbf{x})$ and $\mathbf{b}^1(\mathbf{x})$ vanish on average on the representative volume element. Note that the first order influence functions $^3\mathbf{N}^1(\mathbf{x})$ and $\mathbf{b}^1(\mathbf{x})$ have the physical meaning of periodic micro-fluctuations of the displacement field experienced by the network associated to a variation of the average macroscopic strain and moisture content, respectively.

The second mathematical problem characterizes the average response of the network through the properties of an equivalent homogeneous medium at the macro-scale, i.e. the effective elastic stiffness $^4\overline{\mathbf{C}}$ and hygro-expansion tensor $\overline{\boldsymbol{\beta}}$. They may be expressed as

$$^4\overline{\mathbf{C}} = \frac{1}{|Q|} \int_Q {}^4\mathbf{C}(\mathbf{x}) : (\boldsymbol{\nabla}^3 \mathbf{N}^1(\mathbf{x}) + {}^4\mathbf{I}^S) \mathrm{d}Q \tag{7}$$

$$\overline{\boldsymbol{\beta}} = \frac{1}{|Q|} {}^4\overline{\mathbf{C}}^{-1} : \int_Q {}^4\mathbf{C}(\mathbf{x}) : (\boldsymbol{\beta}(\mathbf{x}) - \boldsymbol{\nabla}\mathbf{b}^1(\mathbf{x})) \mathrm{d}Q \tag{8}$$

To obtain the effective hygro-elastic constants $^4\overline{\mathbf{C}}$ and $\overline{\boldsymbol{\beta}}$, it is required to first solve (in this case using finite elements) the cell problems (5) and (6) for a given network geometry and next to insert their solution into relations (7) and (8). For the derivation of the effective properties, the solution of the cell problems and details on the implementation, the reader may refer to [19].

## 3. Idealized models and their homogenization approaches

The network model and the calculation of its effective hygro-mechanical properties through asymptotic homogenization allow for a detailed representation of the reality, within the simplifications assumed. The procedure used is, however, not trivial and the computational cost to solve the cell problems and to compute the effective constants may be significant. In this Section, two alternative, more idealized representations of the micro-structure of paper are introduced. The first one is based on a lattice description of the fibrous network [20, 21]. The second identifies paper as a perfectly bonded laminate plate, in which all possible orientations are present according to the probability density function. Both models have the advantage that they can be solved analytically, providing estimates of the effective hygro-mechanical properties in a straightforward manner. The corresponding predictions will be compared in Section 4 to the results of the full network model, to assess the accuracy and the range of applicability of the simplified approaches.

### 3.1. Lattice model

The following concepts are based on a recent study [20, 21]. Consider the global reference system $(x, y)$ defined in Section 2. The fibrous network is represented by means of a lattice model with a square unit-cell with edge $L^{lat}$ containing two fibres with orientations, $\theta^{(1)} = 0$ and $\theta^{(2)} = \pi/2$, with respect to the $x-$axis. Each fibre has width $w$ and thickness $t$; a distinction is made between the unbonded, free-standing fibre segment of length $l^{lat} = L^{lat} - w$ and the square bonding area of edge $w$. The orientation distribution is incorporated in the model by integrating the orientation probability density function (1) with respect to $\theta$ at the angles $\theta^{(1)}$, i.e. in the interval $[-\pi/4; +\pi/4]$, and $\theta^{(2)}$, i.e. in the interval $[-\pi/2; -\pi/4] \cup [\pi/4; \pi/2]$. The results of the integration, $\zeta^{(1)}$ and $\zeta^{(2)}$, with $\zeta^{(1)} \geq \zeta^{(2)}$, define the corrected thicknesses of the two fibres oriented at $\theta^{(1)}$ and $\theta^{(2)}$: $t\zeta^{(1)}$ and $t\zeta^{(2)}$ respectively. The bond is considered as a laminated composite plate of thickness $t(\zeta^{(1)} + \zeta^{(2)})$. The geometry of the unit-cell is described in Figure 2(a).

The constitutive behaviour of the fibres and of the bond is given by the same hygro-elastic constitutive laws defined in Section 2.2. A simplification is made in the description of the free-standing segments of the fibres, which are modelled as trusses, i.e. the transverse components of the hygro-elastic deformation are neglected. The bond stiffness $^4\mathbf{C}^b$ and hygro-expansion tensor $\boldsymbol{\beta}^b$ are computed from (4) assuming $n^b = 2$, $\lambda^{(k)} = \zeta^{(k)}/\bar{c}$ and specifying the fibres orientations at $[0, \pi/2]$.

The effective hygro-mechanical properties of the lattice are obtained analytically through homogenization, assuming consistency of the unit-cell strains with the applied macroscopic strain and energy equivalence between the two scales [20, 21]. This allows to recover the effective hygro-mechanical properties as:

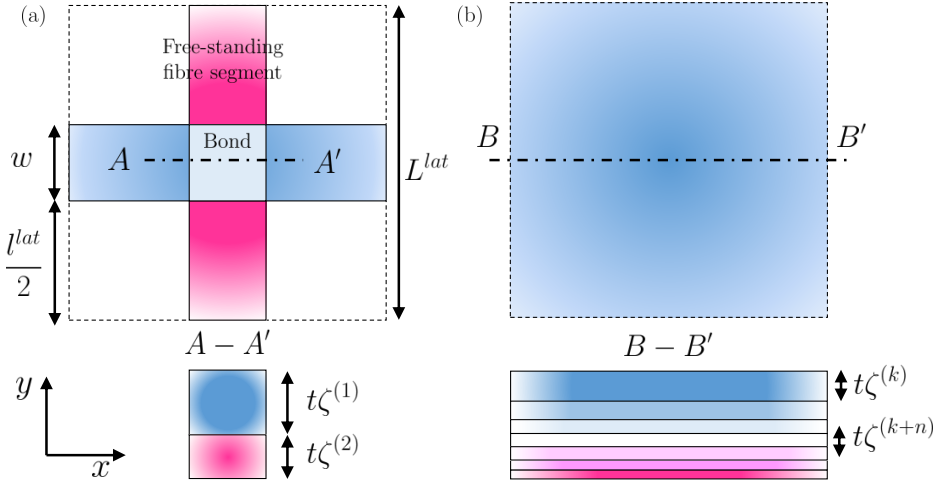

Figure 2: (a) Geometry of the lattice model and bond cross section; (b) Idealized Voigt plate.

$$\overline{E}_{xx}^{lat} = \left[\frac{l^{lat}}{\lambda^{(1)}wE_\ell} + \frac{1}{(C_{xx}^b - \nu_{xy}^b C_{yx}^b)}\right]^{-1} \; ; \; \overline{E}_{yy}^{lat} = \left[\frac{l^{lat}}{\lambda^{(2)}wE_\ell} + \frac{1}{(C_{yy}^b - \nu_{yx}^b C_{xy}^b)}\right]^{-1} \qquad (9)$$

$$\overline{\beta}_{xx}^{lat} = \frac{1}{L^{lat}}(l^{lat}\beta_\ell + w\beta_{xx}^b) \; ; \; \overline{\beta}_{yy}^{lat} = \frac{1}{L^{lat}}(l^{lat}\beta_\ell + w\beta_{yy}^b) \qquad (10)$$

Note that the described unit-cell does not possess any resistance to shear. This affects only minimally the predicted mechanical response of the material in the two principal directions, as well as the hygro-expansion coefficients. To incorporate the shear stiffness, two additional families of fibres can be added to the model, see [20].

To compare the responses of the lattice and network models, the length $L^{lat}$ of the unit-cell edge should be defined as a function of the coverage $\bar{c}$. This can be done by imposing mass equality between the continuum level and the unit-cell, yielding $L^{lat} = n_f^{lat}w/\bar{c}$ where $n_f^{lat} = 2$ is the number of fibre families considered and definition (2) has been used. According to the above expression, the lattice model assumes a limit configuration for $\bar{c} = 2$, when $L^{lat} = w$. In this situation, the unit-cell is entirely covered by the bond. At $\bar{c} = 2$, the effective properties (9) and (10) reach their maximum values, which will be constant for any coverage $\bar{c} \geq 2$.

### 3.2. Voigt average

In the second idealized approach, paper is considered as a composite laminate consisting of an infinite number of layers, whose orientations are defined according to the probability density function $f(\theta)$ given by (1). A sketch of the layered composite plate is shown in Figure 2(b). Each of the layers has hygro-mechanical properties $^4\mathbf{C}^{f(k)}$ and $\boldsymbol{\beta}^{f(k)}$ defined as a function of the orientation $\theta^{(k)}$, recovered from the local axes properties by performing the same axes transformation discussed in Section 2.2. The effective hygro-mechanical properties are computed by extending expressions (4) to integral representations:

$$^4\overline{\mathbf{C}}^V = \int_{-\pi/2}^{\pi/2} f(\theta)^4\mathbf{C}^{f(k)}(\theta)\, \mathrm{d}\theta \qquad (11)$$

$$\overline{\boldsymbol{\beta}}^V = (^4\overline{\mathbf{C}}^V)^{-1} : \int_{-\pi/2}^{\pi/2} f(\theta)^4\mathbf{C}^{f(k)}(\theta) : \boldsymbol{\beta}^{f(k)}(\theta)\, \mathrm{d}\theta \qquad (12)$$

Note that in this case all orientations experience the same total deformation. This is expected to provide the Voigt upper bound for the hygro-mechanical properties, to which paper converges for high coverages (highly dense networks).

## 4. Results and model comparison

### 4.1. Network parameters

The material parameters used to compute the effective properties of paper sheets have been extracted from different studies in the literature. An average fibre length $l = 3.5$mm is assumed [27]; the width of the fibre $w$ has been taken as a function of $l$ through the aspect ratio $l/w = 100$ [1], leading to $w = 35\mu$m. Note that in all the previous derivations only the ratio between the thickness of the individual fibres and the thickness of the bonds plays a role; therefore the value of $t$ does not represent a parameter for the model. The longitudinal elastic modulus is chosen as $E_\ell = 35$GPa [28]. The transverse elastic modulus $E_t$ and the shear modulus $G_{\ell t}$ are defined with respect to the longitudinal stiffness through the constants $\alpha = E_\ell/E_t = 6$ and $\omega = E_\ell/G_{\ell t} = 10$ [29], yielding $E_t = 5.83$GPa and $G_{\ell t} = 3.5$GPa. The Poisson's ratio $\nu_{\ell t}$ is assumed as $\nu_{\ell t} = 0.3$ [29], from which follows $\nu_{t\ell} = \nu_{\ell t}/\alpha = 0.05$. The longitudinal hygro-expansive coefficient $\beta_\ell = 0.013$ is based on [30]. The transverse hygro-expansion is expressed as a function of the longitudinal expansivity as $\beta_t = \gamma\beta_\ell$, with $\gamma = 20$ [1].

The effective material properties are computed by considering several network configurations, corresponding to a set of different coverages $\bar{c} = [0.25, 0.5, 1, 2, 5, 10]$. The minimum size of the representative volume element has been determined by analyzing the effective material properties as a function of the ratio $L/l$ between the size of the unit-cell and the fibre length, for all the considered coverage values. For a cell size of $L \geq 2l$, the value of the effective properties, averaged over the number of network realizations, is within a range of 5% from the properties corresponding to the largest unit-cell size ($L = 8l$), independently from the coverage. For this reason, a cell size of $L = 2l = 7$mm is considered to be representative of the effective material response. For this cell size, the maximum deviation from the effective properties given by an individual network realization from the averaged value over all the realizations is about 8% for coverages $\bar{c} \geq 1$; for larger coverages the deviation is lower than 3%. The number of realizations used can be thus also considered sufficient. For details the reader may refer to [19].

The number of fibres considered, $n_f$, depends on the coverage through relation (2). This results in $n_f = [100, 200, 400, 800, 2000, 4000]$ for $\bar{c} = [0.25, 0.5, 1, 2, 5, 10]$, respectively.

The fibrous networks have been discretized using bilinear quadrilateral plane-stress elements. The elements are all square and have the same size $l_e = w/\xi$. Here, $\xi = 5$ has been used, yielding $l_e = 7\mu$m and a regular grid of 1000×1000 elements.

### 4.2. Isotropic case: uniform fibre orientation

The case of paper with a uniform orientation distribution ($q = 0$) is first analysed. The effective properties are calculated for the set of different coverages $\bar{c}$, considering five network configurations for each value of $\bar{c}$. Figure 3(a) and Figure 3(b) illustrate the orientation-averaged effective elastic moduli $(\overline{E}_{xx} + \overline{E}_{yy})/2$ and hygro-expansion coefficients $(\overline{\beta}_{xx} + \overline{\beta}_{yy})/2$, respectively, as a function of the coverage. The averages $(\overline{E}_{xx} + \overline{E}_{yy})/2$ and $(\overline{\beta}_{xx} + \overline{\beta}_{yy})/2$ have been considered to minimize the randomness of the fibre orientation, which may not yield a perfectly isotropic network. The square markers represent the average values of the five considered realizations of the full micro-structural networks. The error bars illustrate the variation between the different realizations. Both the elastic moduli and the hygro-expansive coefficients increase as a function of the network coverage $\bar{c}$, approaching an asymptotic value. This response is reasonable if one interprets the coverage as a measure of the effective number of fibre layers within a sheet. For increasing layer number (i.e. for a denser network), the degree of internal constraint of the network increases, therefore yielding a larger overall stiffness. Note, finally, that paper properties are generally shown as a function of density rather than of coverage [1]. However, this measure is ambiguous in the proposed approach, due to the impossibility to define the porosity and the actual thickness for a two dimensional system.

The hygro-mechanical properties obtained via asymptotic homogenization of the fibrous networks are compared to those calculated using the idealized models introduced in Section 3. The continuous blue lines represent the estimates from the lattice model according to (9)-(10), in which the length of the unit-cell depends on the coverage via $L^{lat} = 2w/\bar{c}$. The dashed magenta lines indicate the Voigt average (12).

Consider first Figure 3(a) illustrating the mechanical properties. The elastic moduli calculated through the lattice model overestimate (except at $\bar{c} = 1$) the network response. This can be explained by the fact that, whereas the network model and Voigt average are isotropic, the lattice is not. The lattice model has fibres only in $x-$ and $y-$direction, in which they contribute optimally to the stiffness depicted here. To illustrate this, consider the value of the elastic modulus $\overline{E}^{lat}_{\pi/4}$ of the lattice model along an angle $\theta = \pi/4$ with respect to the $x-$axis, indicated in Figure 3(a) at $\bar{c} = 2$ by the blue diamond marker. The Voigt average, to which the network response converges for $\bar{c} \geq 5$, is approximatively between $\overline{E}^{lat}_{\pi/4}$ and $\overline{E}^{lat}_{xx}$. A similar reasoning can be followed for $\bar{c} < 2$: the lattice model yields larger elastic moduli than the network in the $x-$ and $y-$ direction because its response is anisotropic. In this case, however, it is not possible to show the lattice stiffness along $\theta = \pi/4$ as the assumed configuration of the lattice unit-cell does not possess any resistance to shear.

Figure 3(b) shows the effective hygro-expansive properties. The lattice model captures closely the network hygro-expansion for low coverages ($\bar{c} \leq 1$), while for higher coverages ($\bar{c} \geq 5$) the Voigt estimate is quite accurate. In this case the Voigt prediction coincides with the outcome of the lattice model for $\bar{c} \geq 2$. The reason is that, contrary to the stiffness tensor, the hygro-expansion tensor of a two layered bond is isotropic, just like the Voigt hygro-expansion. In general, for both the mechanical and hygro-expansive properties, three regimes can be distinguished from Figure 3: a strong dependence of the effective properties on the coverage for $\bar{c} \leq 1$ (sparse networks), constant properties for $\bar{c} \geq 5$ (dense networks) and a transition zone in between.

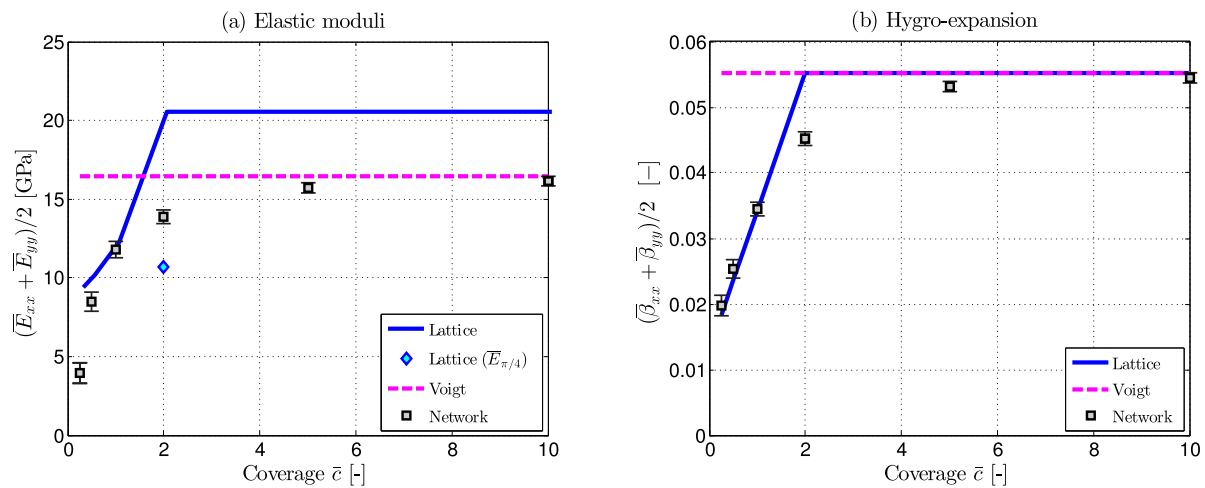

Figure 3: (a) Averaged effective elastic moduli $(\overline{E}_{xx} + \overline{E}_{yy})/2$ and (b) averaged effective hygro-expansion components $(\overline{\beta}_{xx} + \overline{\beta}_{yy})/2$ as a function of the coverage $\bar{c}$, computed with the network, the lattice and the Voigt models.

### 4.3. Anisotropic case: effect of orientation distribution

The effect of the anisotropy of the fibre orientation distribution on the effective hygro-mechanical properties is investigated by considering networks with different values of the parameter $q$: $q = [0, 0.25, 0.5, 0.75]$. Each data point refers to the average of five network realizations. Two values of coverage have been considered, $\bar{c} = 1$ and $\bar{c} = 10$. Figure 4 shows the elastic stiffness components and the expansion coefficients in direction $x$ (magenta markers) and $y$ (blue markers) as a function of the parameter $q$. The variation between the different realizations is shown through the error bars. The effective elastic modulus $\overline{E}_{xx}$ increases as the fibres are more aligned in the machine direction; at the same time, $\overline{E}_{yy}$ decreases. The hygro-expansion coefficients show the opposite tendency: $\overline{\beta}_{yy}$ is larger than $\overline{\beta}_{xx}$, which decreases for increasing $q$, asymptotically approaching (for any coverage) the value of $\beta_\ell = 0.013$. This represents the limit case for $q \to 1$, when all fibres are oriented in the machine direction. The hygro-expansion in the cross direction $\overline{\beta}_{yy}$ is more influenced by the anisotropy and for $q \to 1$, $\overline{\beta}_{yy} \to \beta_t$.

In Figure 4(a) and Figure 4(b), the network's effective response for $\bar{c} = 1$ is compared with the results of the lattice model. Whereas the elastic moduli are generally well captured by the lattice description, the hygro-expansive coefficients can be effectively approximated for low levels of anisotropy only ($q \leq 0.25$). For higher anisotropy, the error increases to 40%, especially in the cross direction. Note that the Voigt solution is not shown here because it is approached by the network solution only for higher coverage values, $\bar{c} \geq 5$. Figures 4(c) and Figure 4(d) compare the network effective properties for $\bar{c} = 10$ to the lattice (continuous lines) and Voigt (dashed lines) estimates. As in the isotropic case, the lattice overestimates the elastic moduli of the network model, due the former's anisotropy, while the Voigt average offers an adequate approximation. The hygro-expansive coefficients match well with the Voigt estimate, but the even simpler lattice model (i.e. a two layers plate) provides an equally accurate approximation.

The previous observations can be generalized by studying the effect of anisotropy as a function of the coverage $\bar{c}$. Figure 5 shows the hygro-mechanical properties as a function of $\bar{c}$ for $q = 0.25$. For increasing coverage, the stiffness and the hygro-expansion increase in both the machine and the cross direction. The mechanical response is, like for the isotropic case, overestimated by the lattice model. On the other hand, for low coverages, $\bar{c} \leq 1$, it still provides a reasonable approximation for the hygro-expansion coefficients, for the considered low level of anisotropy ($q = 0.25$). For high coverages $\bar{c} \geq 5$, the Voigt bound is effective in capturing both the mechanical and hygroscopic properties. In this range of coverage, the lattice model can also be used to estimate the hygro-expansive response.

Note that, in general, typical values for in-plane elastic properties of paper range from 1 to 10 GPa [31]. The obtained network results in terms of mechanical properties exceed this range. The hygro-expansive properties, however, are very realistic, see e.g. [8], and so are the trends of both elastic and hygro-expansive quantities as a function of coverage and anisotropy of the orientation distribution. The quantative overprediction of the elastic stiffness may be due to the two-dimensional character of the network. The simplified models use that same assumption and hence the network model is a good reference to compare them against. In conclusion, Table 1 presents a summary that defines the range of applicability of each of the proposed models, taking the network model as the reference. For mechanical properties, the network description has to be used, except for high coverages ($\bar{c} \geq 5$), where the Voigt model is a valid alternative. For hygro-expansive properties, at low levels of coverage and for a low degree of anisotropy, the lattice model captures well the regime predicted by the network simulations. For high coverages, for any level of anisotropy, both the Voigt and the lattice model can be used as they adequately approximate the network response. In all other cases, the network model should be preferred.

### 4.4. Parametric study

The presented results have been obtained for the set of micro-structural parameters indicated in Section 4.1. Their values have been taken form the literature; however, it is well known that there is a large variability in material properties of paper [1]. The influence of the choice of the micro-scale parameters on the effective hygroscopic properties of paper is investigated in Figure 6. A reference value of the coverage is chosen as $\bar{c} = 4$, i.e. a network with on average four fibre layers. In this range (see Table 1), the network model should be used to predict the hygro-mechanical properties of paper. In order to observe at the same time the influence of the mechanical and hygro-expansive properties, the hygro-expansive coefficients $\overline{\beta}_{xx}, \overline{\beta}_{yy}$ are shown as a function of the effective stiffness ratio $\overline{E}_{xx}/\overline{E}_{yy}$. The coefficients $\overline{\beta}_{xx}$ and $\overline{\beta}_{yy}$ obtained using the reference parameters ($\alpha = 6, \gamma = 20, l/w = 100, \bar{c} = 4$) are shown in all diagrams of Figure 6 through blue lines, while the variation with respect to the reference configuration is presented in magenta lines. Solid and dashed lines refer to the $x-$ and $y-$directions, respectively.

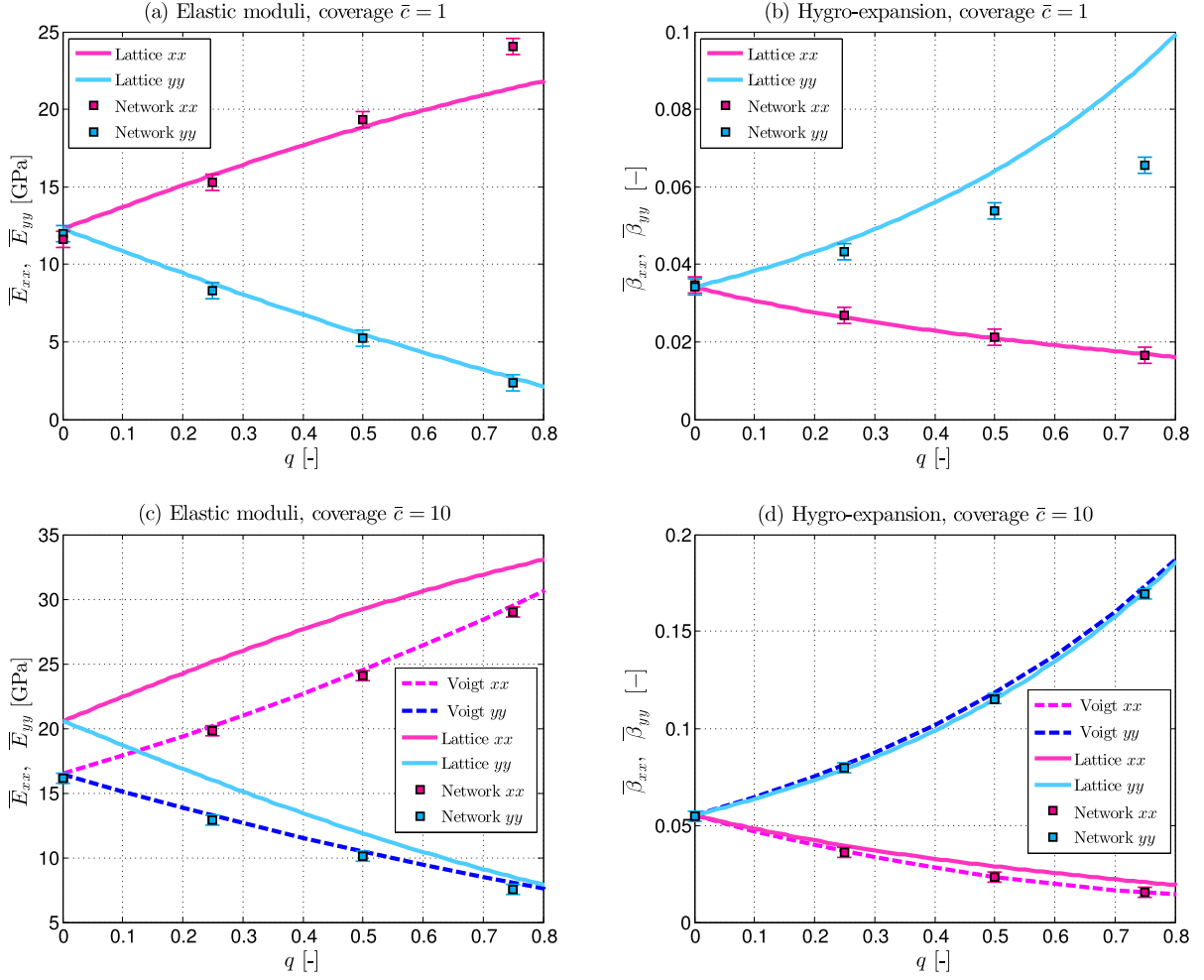

Figure 4: (a)-(c) Effective elastic moduli $\overline{E}_{xx}, \overline{E}_{yy}$ and (b)-(d) hygro-expansion coefficients $\overline{\beta}_{xx}, \overline{\beta}_{yy}$ as a function of the anisotropy parameter $q$ for coverages (a)-(b) $\bar{c} = 1$ and (c)-(d) $\bar{c} = 10$.

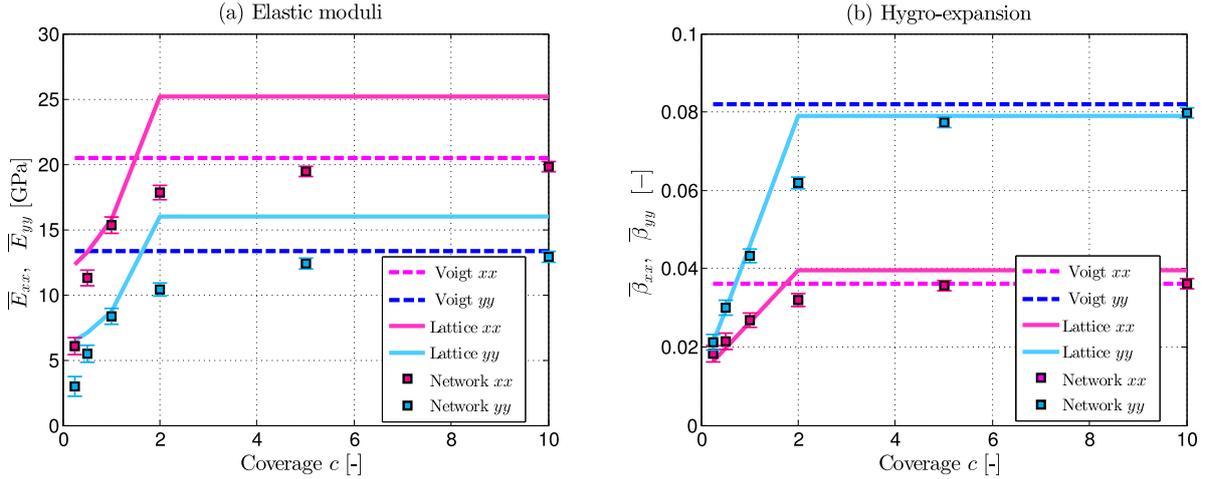

Figure 5: Effective hygro-mechanical properties as a function of the coverage for an anisotropic fibre orientation distribution with $q = 0.25$: (a) elastic moduli $\overline{E}_{xx}, \overline{E}_{yy}$ and (b) effective hygro-expansion coefficients $\overline{\beta}_{xx}, \overline{\beta}_{yy}$.

| Mechanical properties | | | |
|---|---|---|---|
| Coverage | Network | Lattice | Voigt |
| $\bar{c} \leq 1$ | ✓ | | |
| $1 < \bar{c} < 5$ | ✓ | | |
| $\bar{c} \geq 5$ | ✓ | | ✓ |

| Hygro-expansive properties | | | |
|---|---|---|---|
| Coverage | Network | Lattice | Voigt |
| $\bar{c} \leq 1$ | ✓ | For $q \leq 0.25$ | |
| $1 < \bar{c} < 5$ | ✓ | | |
| $\bar{c} \geq 5$ | ✓ | ✓ | ✓ |

Table 1: Range of applicability of each of the proposed models as a function of the network coverage.

Figure 6(a) illustrates the effect of the choice of the ratio $\gamma = \beta_t/\beta_\ell$, between the transverse and the longitudinal fibre hygro-expansion. The case of $\gamma = 10$, also reported in the literature [25], is considered. This parameter has a strong effect not only on the magnitude of the coefficients, but also on the anisotropy of the overall response, reducing the difference between the expansion in machine and in cross direction.

The second parameter investigated is the ratio between the elastic moduli, $\alpha = E_\ell/E_t$. This is shown in Figure 6(b), which presents a case in which $\alpha = 9$ [29]. The change of this parameter does not affect much the magnitude of the hygro-expansion, but, as expected, it decreases the level of anisotropy in the mechanical properties, shown by the larger values assumed by the ratio $\overline{E}_{xx}/\overline{E}_{yy}$.

In Figure 6(c), the effect of the coverage is studied, taking $\bar{c} = 3.8$. This parameter has a large effect on the results; a lower coverage corresponds to lower hygro-expansive properties and a higher ratio between mechanical properties, as already illustrated in Figure 5.

Finally, Figure 6(d), shows that the effect of the fibre aspect ratio, here taken as $l/w = 50$, is limited. This can be explained by the fact that, the coverage being constant, the average bonding area, which greatly influences the effective hygro-mechanical response, does not change.

4.5. Comparison with experiments

The results of the network model are finally compared with the hygro-expansion coefficients determined experimentally in [11]. The experimental data are referred to a paper of density $\rho_p = 420 kg/m^2$. The density relates to the coverage through relation $\bar{c} = b_p/b_f = \rho_p h_p/b_f$, where $h_p$ is the macroscopic thickness of paper and relation (2) has been used. Given the uncertainty in these parameters in the experiments, some assumptions need to be made: i) $h_p = 100\mu$m [21] and $b_f = 10$ g/m$^2$ [1], yielding $\bar{c} = 4.2$, ii) $\alpha = E_\ell/E_t = 10$ [29], iii) $\gamma = \beta_\ell/\beta_t = 7$ [25] and iv) $\beta_\ell = 0.03$ [1]. The present comparison thus has the main goal to show that, for micro-structural parameters that fall in a realistic range, the experimental data can be semi-quantitatively captured by the model. The results of the experiment are represented in Figure 7 with grey lines (dashed in the cross direction, solid in the machine direction) as a function of the effective stiffness ratio $\overline{E}_{xx}/\overline{E}_{yy}$. The corresponding predictions of the network model are illustrated through blue lines. Although they do not capture exactly the experimental data, the trend resembles the experimental one and the resulting values are meaningful, considering the number of unknowns characterizing the problem and the fact that no fitting of the meso-scale parameters has been performed.

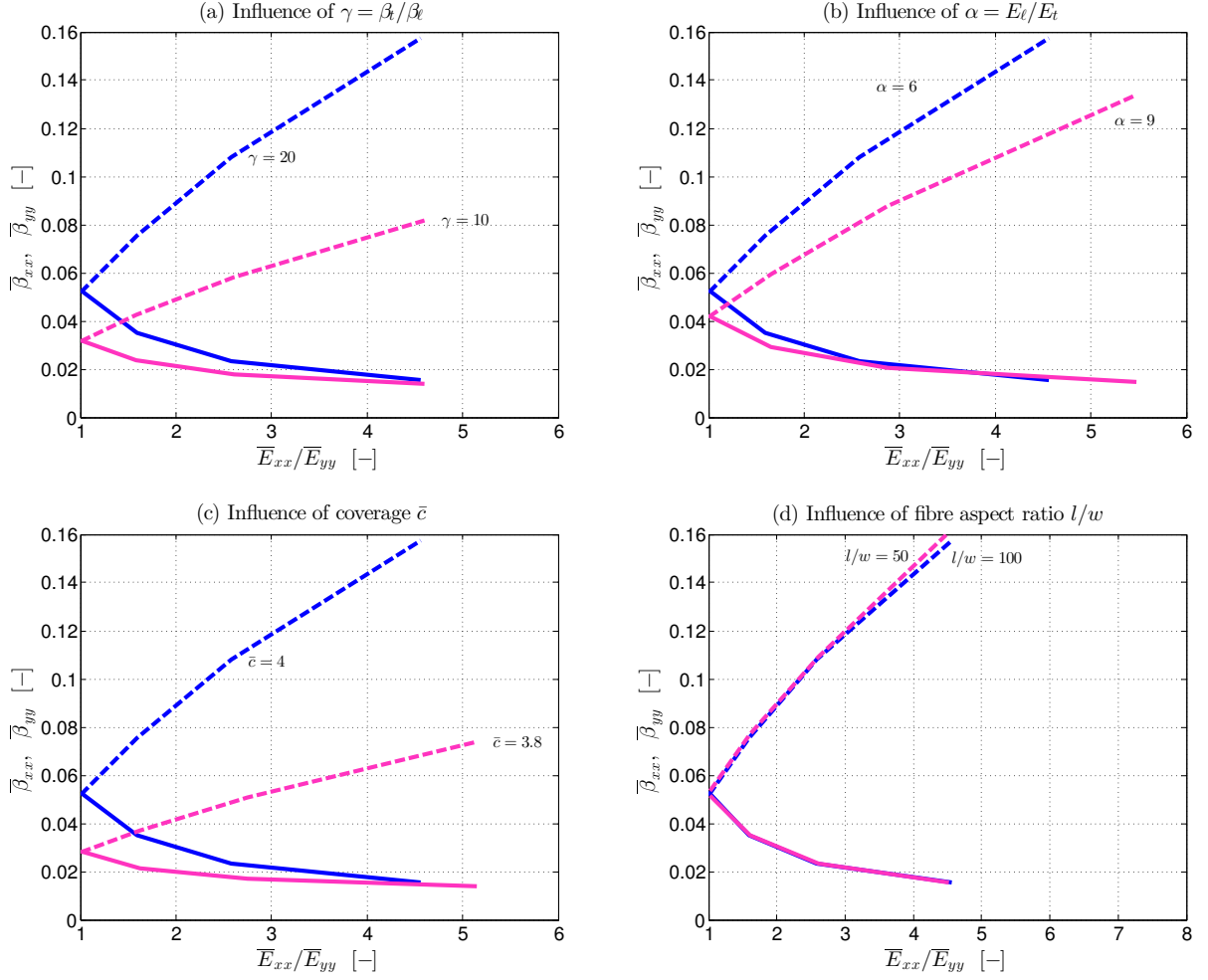

Figure 6: Hygro-expansive coefficients $\overline{\beta}_{xx}, \overline{\beta}_{yy}$ as a function of the effective stiffness ratio $\overline{E}_{xx}/\overline{E}_{yy}$ and influence of different meso-structural parameters: (a) $\gamma = \beta_t/\beta_\ell$, (b) $\alpha = E_\ell/E_t$, (c) coverage $\bar{c}$ and (d) aspect ratio $l/w$.

## 5. Conclusions

This work presents a multi-scale methodology to calculate the effective hygro-mechanical response of paper. To this aim, two distinct strategies have been followed. First, a detailed micro-structural model has been developed, which consists of a random fibrous network characterized by an orientation distribution function. A rigorous asymptotic homogenization approach has been used to compute the effective elastic stiffness and the hygro-expansion tensor, as a function of the micro-structural parameters (hygro-mechanical properties of the single fibres, fibres and network geometry, areal coverage, etc.). This accurate, though computational expensive procedure has been compared with two alternative models of the paper micro-structure. They are based on a highly idealized representation of the underlying network, either using a lattice model consisting of two orthogonal fibres or exploiting the Voigt estimate. The effective hygro-mechanical response is obtained in these cases by analytical homogenization.

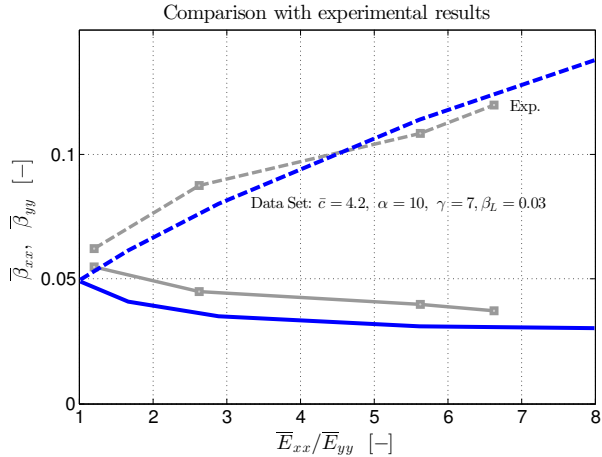

Figure 7: Comparison between the hygro-expansive coefficients $\overline{\beta}_{xx}, \overline{\beta}_{yy}$ predicted by the network model (blue lines) and the experimental data taken from [11] (grey lines).

The results of all the proposed models have been analysed, revealing that the simplified models can be used for practical purposes as good alternatives to the network description. The major conclusions are:

- For values of coverage which are to be expected in paper ($\bar{c} \geq 5$), the idealized lattice model and the Voigt average both capture in an accurate way the effective hygro-expansive coefficients predicted by the network model. The Voigt average correctly predicts also the mechanical response.

- The above considerations hold independently from the anisotropic orientation distribution.

- In the low coverage regime ($\bar{c} \leq 1$) and for poorly oriented networks ($q \leq 0.25$), the lattice model approximates well the effective network hygro-expansion.

- The recovered effective properties capture in a semi-quantitative manner the trend of experimental data.

A more accurate experimental comparison, in which the material parameters are known both at the micro and macroscopic level, would greatly increase the value of this study. This aspect will be tackled in subsequent publications.

Further extensions to this contribution, focusing on the reversible hygro-mechanical response only, can be aimed at incorporating the effects of the irreversible hygro-expansive deformation, along the lines of [32].

Moreover, the work could be developed towards a three dimensional description. This would allow to represent the geometry more accurately and to relax the assumption of full kinematic compatibility in the bonding regions through-the-thickness.

Finally, the effective hygro-mechanical properties obtained from the proposed methodology could be used as input parameters for macro-scale models (e.g. [33, 34]) predicting paper deformations under complex moisture and mechanical loads as required in a number of industrial applications.

**Acknowledgements**

This research was carried out under project number M61.2.12458 in the framework of the Research Program of the Materials innovation institute M2i (www.m2i.nl).